\documentclass[graybox]{svmult}

\usepackage{amsmath} 
\usepackage{seqsplit}
\usepackage{subfig}

\usepackage{mathptmx}       
\usepackage{helvet}         
\usepackage{courier}        
\usepackage{type1cm}        
\usepackage{makeidx}         
\usepackage{graphicx}        
\usepackage{multicol}        
\usepackage[bottom]{footmisc}

\makeindex             

\begin{document}

\title{The effects of amplification of fluctuation energy scale by quantum measurement choice on quantum chaotic systems: Semiclassical analysis}

\author{S. Greenfield$^{1,2,3}$ Y. Shi$^{1}$, J.K. Eastman$^{2,3,4}$, A.R.R.~Carvalho$^{3}$, and A.K. Pattanayak$^{1}$}

\institute{ \at (1) Department of Physics and Astronomy, Carleton College, 1 N. College Street, Northfield, MN 55057, United States of America,
\at (2) Centre for Quantum Computation and Communication Technology (Australian Research Council), 
\at (3) Centre for Quantum Dynamics, Griffith University, Brisbane, Queensland 4111, Australia
\at (4) Department of Quantum Science, Australian National University, Canberra, Australia, 
\at \email{greenfields@carleton.edu,
shiy@carleton.edu,
jessica.eastman@anu.edu.au,
a.carvalho@griffith.edu.au,
arjendu@carleton.edu} 
}
\maketitle 
\abstract{
Measurement choices in weakly-measured open quantum systems can affect quantum trajectory chaos. We consider this scenario semi-classically and show that measurement acts as nonlinear generalized fluctuation and dissipation forces. These can alter effective dissipation in the quantum spread variables and hence
change the dynamics, such that measurement choices on the coupled quantum dynamics can enhance quantum effects and make the dynamics chaotic, for example. This analysis explains the measurement-dependence of quantum chaos at a variety of parameter settings, and in particular we demonstrate quantitatively that the choice of monitoring scheme can be more relevant than system scale $\beta$ in determining the `quantumness' of the system. 
}

\section{Introduction}
Measuring a quantum system has an unavoidable effect in its state. This is a feature with no classical counterpart that introduces an entirely quantum pathway to manipulate quantum systems. In particular, the continuous monitoring of a quantum system provides the ability to implement real-time control, which can be used to enhance or suppress desirable effects in the system dynamics.  
Recent work has shown that continuously measured open quantum system trajectory dynamics can change between the qualitatively dramatic different regimes 
of chaos (with high dynamical algorithmic complexity) and regularity (with qualitatively different dynamical complexity) depending on parameter choices~\cite{Eastman:2017,Pokharel:2018}. 
In particular, the phase $\phi$ setting on a laser used as the local oscillator for making a homodyne measurement of the signal from a driven dissipative nonlinear quantum oscillator were shown to considerably affect the system dynamics \cite{Eastman:2017}. 
The back-action from this kind of measurement manifests as a generalized dissipation $\vec{F}(\phi)$ and `noise' $\vec{N}(\phi)$ where changes in $\phi$ can 
strongly affect the quantum dynamics sometimes making them chaotic, depending in a puzzling way on a combination of system parameters, 
including size, and the behavior of the classical limit. Understanding this puzzle would help us use $\phi$, an external experimentally accessible parameter, to control quantum trajectories in useful ways.

We consider this system in the semi-classical regime where the measurement localization allows us to accurately and efficiently simulate the quantum state as a wave packet described completely by the coupled dynamics of its expectation values (centroid) and variances (spread). 
We use a formalism\cite{Pattanayak:1997} representing $|\psi(t)\rangle$ as the dynamics of two oscillators: the centroid $(x,p)$ and the spread $(\chi,\Pi)$ of the wave packet (detailed definitions below). Without environmental coupling these evolve according to the Hamiltonian $H(x,p,\chi,\Pi) = p^2/2 + \Pi^2/2 + U(x,\chi) = H_1(x,p) + H_2(\chi,\Pi) + U_{12}(x,\chi)$ where the size of the `quantum' Hamiltonian $H_2$ and the coupling $U_{12}$ increases relative to $H_1$ as the system becomes smaller in size, say leading to an increase in the effective Planck's constant $\beta$.
Thus, the influence of the quantum oscillator on the classical motion increases with $\beta$ via $U_{12}(x,\chi)$. The environment acts with $\vec{N}$ coupling only to $(x,p)$ and the $\phi-$dependent part of $\vec{F}$ coupling only to $(\chi,\pi)$. 
Energy analysis is useful to understand the non-trivial effect of changing $\vec{N}$ and $\vec{F}$ with $\phi$. Small changes in the  fluctuation and dissipation $\vec{N},\vec{F}$ change how the nonlinear dynamics amplify the quantum fluctuations and significantly change the energy range for the dynamics for $\chi,\Pi$. 
This change alters the $U_{12}(x,\chi)$ coupling and hence the influence of the quantum oscillator on the classical dynamics. 

We consider several such $(\Gamma,\beta,\phi)$ combinations to consider the effects of changing these parameters on the various competing effects. 
Our simulations verify our energy-based explanation for $\phi$-dependent quantum trajectory chaos.
We also find that measurement angle $\phi$ can affect the relative quantum energy scale compared to classical one by orders of magnitude more than the system scale $\beta$.

Below, we review the coupled-oscillator formalism then focus on the $\phi$ dependence of $\vec{F}$ and $\vec{N}$ before
presenting our results and analysis. We conclude with a discussion about adaptive control of quantum trajectories as well as prospects for experimental implementations of these ideas.

\section{Semi-classical coupled oscillator model} 
Our analysis starts with the quantum model of a damped driven Duffing oscillator~\cite{Brun:1996,Kapulkin:2008,Ota:2005,Eastman:2017,Pokharel:2018}. The Hamiltonian 
$\hat{H}_{D} = \frac{1}{2} \hat{P}^{2}+\frac{\beta^{2}}{4}\hat{Q}^{4} -\frac{1}{2} \hat{Q}^{2} - \frac{g}{\beta} \hat{Q} \cos(\omega t)$ describes the double-well oscillator driven sinusoidally with strength $g$ in terms of dimensionless position ($\hat Q$) and momentum ($\hat P$) operators. $\beta$ serves as a dimensionless effective Planck's constant~\cite{Brun:1996,Ota:2005}: larger $\beta$ describe a smaller system and $\beta\to 0$ is the classical limit. Quantum mechanical damping is introduced via the interaction of the system with a zero-temperature Markovian bath, which corresponds to having $\hat a=(\hat Q + i \hat P)/\sqrt{2}$ in the decoherence superoperator~\cite{Gorini:1976, Lindblad:1976}. Furthermore, we consider that this dissipative quantum channel is being weakly and continuously monitored, such that the state of the system evolves conditioned on the measurement outcomes as given by the following Ito stochastic equation~\cite{Wiseman:2001b,Rigo:1996}
\begin{eqnarray}
	|d\psi\rangle &=& \left(-\frac{i}{\hbar}\hat H + \langle \hat L^\dagger \rangle \hat L
	-\frac{\hat L^\dagger \hat L}{2}
	-\frac{\langle \hat L^\dagger \rangle \langle \hat L\rangle}{2} \right) |\psi\rangle dt 
    + (\hat L - \langle \hat L\rangle )|\psi \rangle
	d\xi.
	\label{sse}
\end{eqnarray}
\noindent Here, $\hat{L} = \sqrt{2\Gamma} \hat a$ represents the dissipative environmental interaction of strength $\Gamma$, and $\hat{H}=\hat{H}_{D}+\hat{H}_{R}$. Since the quantum dissipation is symmetric in $\hat Q$ and $\hat P$, the term $\hat{H}_{R} = \frac{\Gamma}{2} \left(\hat{Q} \hat{P} + \hat{P}\hat{Q} \right)$ is added to yield the correct classical limit where dissipation appears only in the momentum variable.
The noisy dynamics is given in terms of a complex-valued Wiener process, $d\xi$, with $M(d\xi) =0, M (d\xi d\xi^*)=dt$, and $M (d\xi d\xi)= u \, dt$, where $M(\cdot)$ denotes the mean over realizations and the complex parameter $u=\vert u \vert e^{-2 i \phi}$ must satisfy the condition $\vert u\vert \le 1$~\cite{Wiseman:2001b,Rigo:1996}. Here we will consider the situation where $\vert u\vert=1$, which has been shown to correspond to monitoring the dissipative channel with a quantum optical homodyne measurement~\cite{Wiseman:2001b,Eastman:2017} with $\phi$ being the phase of the local oscillator. In this case, the noise can be written as $d \xi= e^{-i\phi} dW$, where $dW$ is a real Wiener process. Recent analysis\cite{Li:2012} shows that nano-electro-mechanical systems are well described by this model and current experiments are within range of the phenomena we report.

A semi-classical analysis starting with the dynamics of $\langle\hat Q\rangle = x, \langle\hat P\rangle = p$ proves
very useful\cite{Halliwell:1995,Ota:2005,Pokharel:2018}; the centroid variables' dynamics depend on second moment terms 
$V_{QQ},V_{PP},V_{PQ}$ where $V_{AB} = \langle (\hat A^\dagger - \langle\hat A\rangle^*) (\hat B - \langle\hat B\rangle) \rangle$. 
In this limit, $|\psi(t)\rangle$ is accurately and completely described by the 
$4D$ phase-space vector $\vec{X} = (x,p,\chi,\Pi)$ with dynamics given by
\begin{subequations}
\label{eq:full}
\begin{align}
\dot{x} &= p + \sqrt{\Gamma}  N_x(\phi,\chi,\Pi) \, dW,
\label{eq:xdot}\\
 \dot{p} &=  x -\beta^2 x^3 +\frac{g}{\beta}\cos{\omega t} +\Gamma F_p + 3 x\beta^2\chi^2 + \sqrt{\Gamma}  N_p(\phi,\chi,\Pi) \, dW\label{eq:pdot},\\
\dot{\chi} &= \Pi + \Gamma F_{\chi}(\phi,\chi,\Pi)\label{eq:chidot},\\
\dot{\Pi} &= \chi(-3\beta^2 (x^2 + \chi^2) + 1) 
+ \frac{1}{4 \chi^3} + \Gamma F_{\Pi}(\phi,\chi,\Pi) \label{eq:pidot}, 
\end{align}
\end{subequations}
where the change of variables 
$V_x = \chi^2,
V_{xp} = \chi \Pi,
V_{p} = 1/4\chi^2 + \Pi^2
$
is for convenience below. The random effect of the continuous monitoring is given by the stochastic terms $\vec{N} = (N_x,N_p,N_\chi,N_\Pi)$ with 
\begin{subequations}
\begin{align}
N_x &= 2(\chi^2 - \frac{1}{2}) \cos{}(\phi) - 2\chi \Pi \sin{}(\phi),\\ 
N_p &= -2(\frac{1}{4 \chi^2} + \Pi^2 - \frac{1}{2})\sin(\phi) + 2\chi \Pi \cos(\phi),
\end{align}
\end{subequations}
while $N_\chi = 0 = N_\Pi$. The dissipation $\vec{F} = (F_x,F_p,F_\chi,F_\Pi)$ has $F_x = 0, F_p = -2\Gamma$ 
and  
\begin{subequations}
\label{eq:Fdiss}
\begin{align}
\label{eq:Fchi}
F_\chi &=  
\bigg [\chi -\chi^3 + \chi \Pi^2 - \frac{1}{4 \chi} \bigg ] \cos(2\phi) 
-\Pi \bigg [ -1 + 2\chi^2 \bigg ]\sin(2\phi) \nonumber\\ 
&+\chi - \chi^3 - \chi \Pi^2 + \frac{1}{4 \chi},\\
F_\Pi &= \bigg [\Pi^3 -\Pi+
 \frac{3\Pi}{4\chi^2}-\Pi\chi^2\bigg ]\cos(2\phi)
 + \bigg [-\frac{1}{4\chi^3}+\frac{1}{\chi}-\chi+2\chi\Pi^2\bigg ]\sin(2\phi)
\nonumber\\
 &+
 (-\Pi^3 -\Pi-\frac{3\Pi}{4\chi^2}-\Pi\chi^2). \label{eq:Fpi}
\end{align}
\end{subequations}

\section{Coupling between centroid and spread oscillators}

With the model from the previous section, we can now describe how the spread oscillator, given by the canonically conjugate pair $(\chi,\Pi)$, influences the dynamics of the classical oscillator, given by the centroid variables $(x,p)$.

\begin{figure}
 \centering
    \subfloat[$\beta = 0.05$]{{\includegraphics[width=5cm]{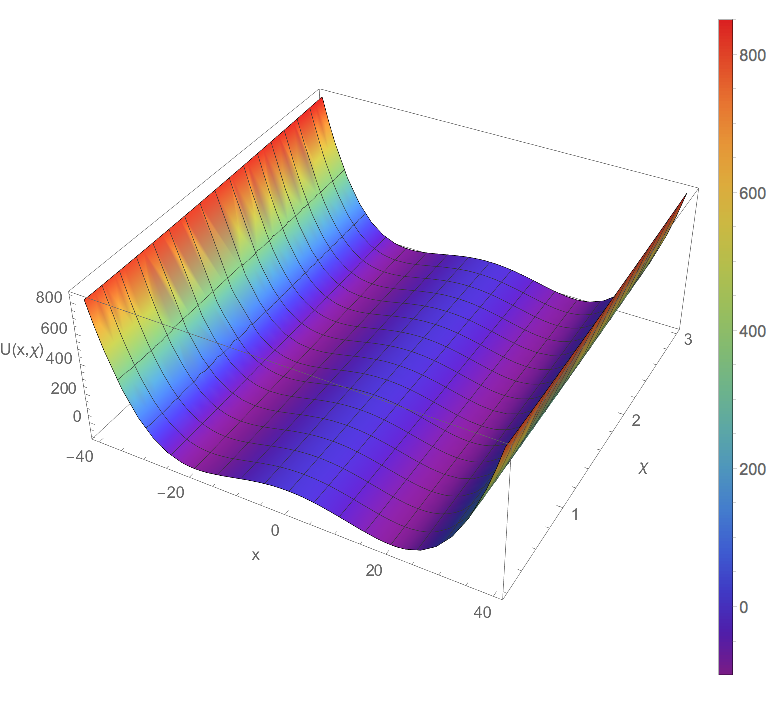} }}%
    \qquad
    \subfloat[$\beta = 0.3$]{{\includegraphics[width=5cm]{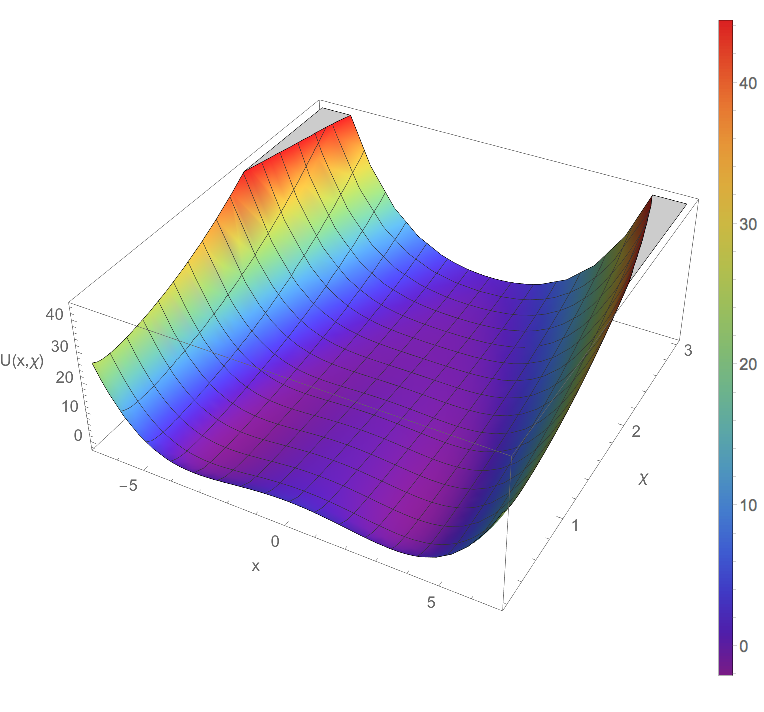} }}%
    \caption{Potential $U(x,\chi,t)$ for $g=0$ and (a) $\beta = 0.05$, and (b) $\beta = 0.3$. For larger $\beta$, a path between the two wells is possible through higher values of $\chi$.}%
    \label{fig:Potential}%
\end{figure}

For $\Gamma \rightarrow 0$, equations~(\ref{eq:full}) have a Hamiltonian structure with
\begin{eqnarray}
\label{eq:H}
H(x,p,\chi,\Pi) &=& \frac{1}{2}p^2+ \frac{1}{2} \Pi^2 
+ U(x,\chi,t).
\end{eqnarray} Thus, we can represent $\vec{X}(t)$ as a point trajectory traveling in a time-dependent $2D$ semi-classical potential, $U(x,\chi,t)=U_1(x,t) + U_2(\chi) + U_{12}(x,\chi)$, given in terms of 
\begin{eqnarray}
U_1(x,t) &=& -\frac{1}{2} x^2 + \frac{1}{4}\beta^2 x^4 +\frac{g}{\beta}x\cos{\omega t},\\
U_2(\chi) &=& \frac{3}{4}\beta^2\chi^4 - \frac{1}{2}\chi^2 + \frac{1}{8 \chi^2},\\
U_{12}(x,\chi)&=& \frac{3}{2}\beta^2 x^2 \chi^2.
\end{eqnarray} 
The $U(x,\chi,t)$ potential is shown in Fig.~\ref{fig:Potential} for $g=0$ and two different values of $\beta$. The driving sinusoidally tilts the potential along $x$, rocking the particle between the two classical wells depending on the amplitude. 

The inter-oscillator coupling $U_{12}$, which allows the classical and quantum oscillators to influence each other, only exists for nonlinear systems. 
Different dynamical regimes can be quantified via the relative $\beta$ dependence of $\overline{U}_1,\overline{U}_2,\overline{U}_{12}$ where the overbar represents a time average over the trajectory: 
\begin{itemize}
\item For $\beta\to 0, U_{12} \to 0$ and the quantum $(\chi,\Pi)$ dynamics do not influence the classical $(x,p)$ dynamics. These latter are invariant\cite{Brun:1996,Ota:2005,Kapulkin:2008} under change of $\beta$. That is, the phase-space dynamics are identical except that the scale of $x$ increases as $\beta^{-1}$, and $U_1 \sim \beta^{-2}$.
\item The near-classical limit $\beta \ll 1$ has $\overline{U}_1 >> \overline{U}_{12} \simeq \overline{U}_2$. In Fig.~(\ref{fig:Potential}) at $\beta = 0.05$ we can see that this results in a well where 
the classical double-well shape is seemingly barely altered by quantum effects in the typical dynamical range for $\chi$, which is natural since $\overline{U}_1 >> \overline{U}_{12}$. 
\item As $\beta$ increases, we get that $\overline{U}_1 \ge \overline{U}_{12} \simeq \overline{U}_2$. We see in Fig.~(\ref{fig:Potential}) that for $\beta=0.3$ this changes $U(x,\chi)$ in the $\chi$ direction, and creates a non-classical path from one $x$ well minimum to the other that avoids the well maximum at increased $\chi$, considerably altering the dynamics for $(x,p)$ in the process. 
\end{itemize}
This $\beta$ regime where $\overline{U}_1 \geq
\overline{U}_{12} \simeq \overline{U}_2$ is our focus.
When $\overline{U}_1 \simeq
\overline{U}_{12} \simeq \overline{U}_2$ we expect quantum effects to matter in a way that is not visible in semi-classical dynamics.
It is important to realize that systems dynamics and dissipation can alter $U_{12}$ dramatically. In particular, the time dependence of quantum spread variables depends on the components of the Jacobian of classical dynamics. That is, not only does the $U_{12}$ coupling between the two oscillators only exist for nonlinear systems, but as $(\chi,\pi)$ is being dragged around by $(x,p)$ in this regime, the same dynamical properties that cause the chaotic separations of $(x,p)$ trajectories in time causes the $(\chi,\pi)$ spread oscillators to grow and oscillate more rapidly; that is, chaotic dynamics can nonlinearly amplify $U_{12}$ in principle. The constraining factor is the dissipation, as we see below.

\section{Measurement-dependent dissipative forces and oscillator energetics} 

\begin{figure}[hb]
 \centering
    \subfloat[$F_c$]{{\includegraphics[width=5cm]{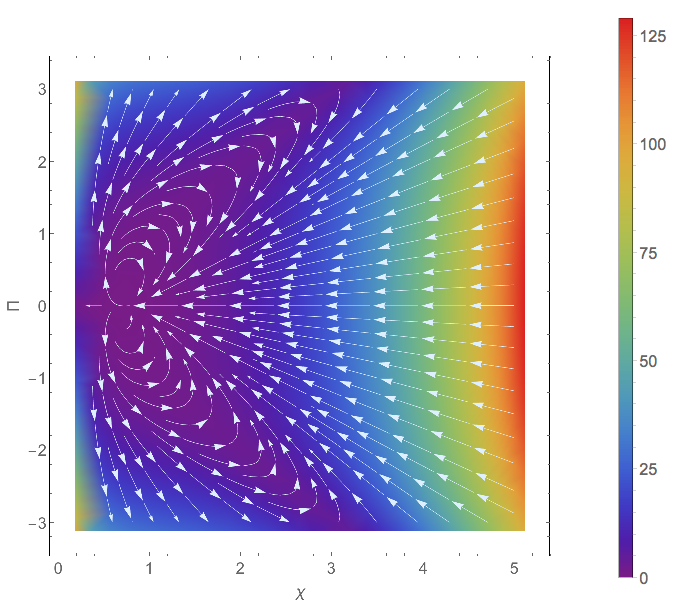} }}%
    \qquad
    \subfloat[$F_s$]{{\includegraphics[width=5cm]{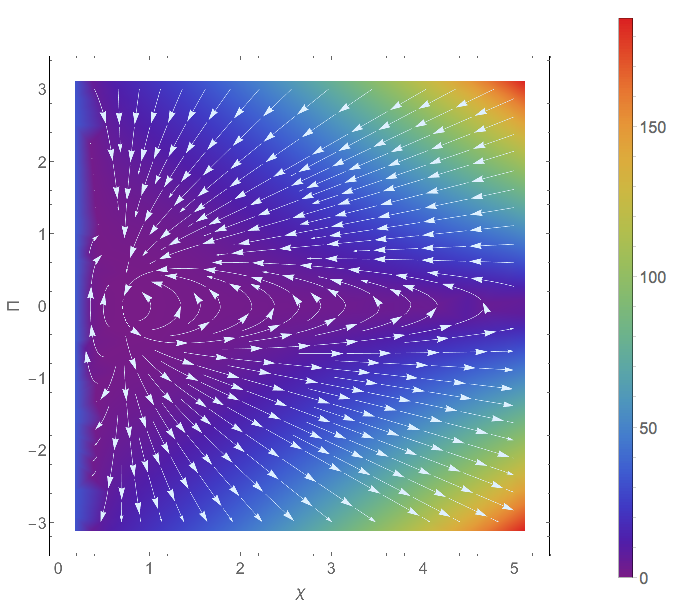} }}%
    \qquad
    \subfloat[$F_0$]{{\includegraphics[width=5cm]{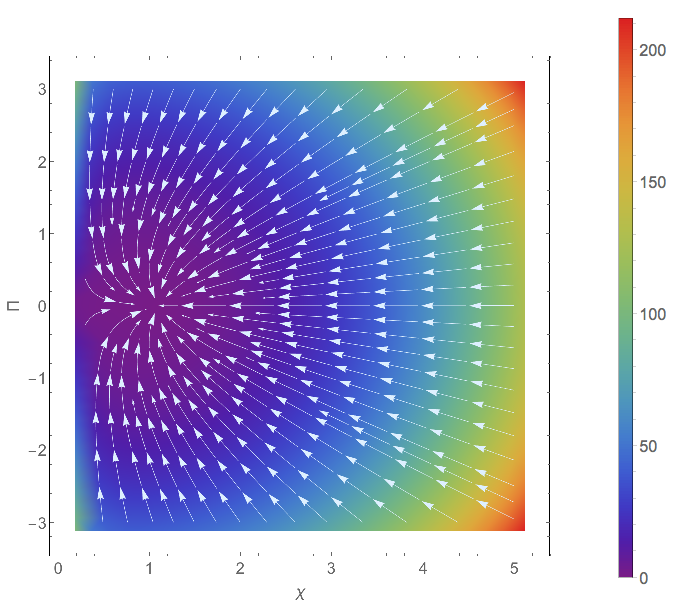} }}%
    \caption{Plots depicting the magnitude and direction of $\vec{F}_0,\vec{F}_c,\vec{F}_s$.Different measurement angles correspond to a weighted superposition. The differences in the $\vec{F}_c,\vec{F}_s$ components pushes the $(\chi,\Pi)$ orbit to different scales, changing the coupling to the classical $(x,p)$ oscillator.}
\label{fig:FPlots}%
\end{figure}

To see how the measurement angle $\phi$ affects the dynamics, we rewrite the dissipative forces as $\vec{F} = (F_x,F_p,F_\chi,F_\Pi) = \vec{F_c}\cos{2\phi} + \vec{F_s}\sin{2\phi} + \vec{F_0}$, where the definitions of $\vec{F_c},\vec{F_s},\vec{F_0}$ are evident from the form of Eqs.(\ref{eq:Fdiss}). 
Defining these three components, which are shown in Fig.~(\ref{fig:FPlots}), is useful since all $\vec{F}$ are weighted superpositions of them. In particular, at $\phi = 0$, $\vec{F} = \vec{F_0} + \vec{F_c}$ and at $\phi = \pi/2, \vec{F} = \vec{F_0} - \vec{F_c}$. In the latter, the contributions of $\vec{F_0}$ and $\vec{F_c}$ along the $\Pi=0$ axis are in opposite directions and tend to cancel out, while in the former, they add up, forcing the system towards small values of $\chi$. 
Note that, in this case, by suppressing higher $\chi$ values, the dissipative force works against the non-classical mechanism for inter-well transitions explained in the previous section. In either case, while the size of the $\Gamma$ governs how the 
driving energy absorbed is dissipated, it is the measurement angle $\phi$ that effectively alters the energy flow between the two oscillators.

To make the connection with energy flow more evident, we can look at how the input power, introduced by the external driving term, is distributed over the different available channels. From conservation of energy, we can write that
\begin{equation}
\frac{dE_g(\vec{X}(t))}{dt} + 
\frac{dE_{\Gamma}(\vec{X}(t))}{dt} + 
\frac{dE_{\sqrt{\Gamma}}(\vec{X}(t))}{dt} + 
\frac{dE_{H}(\vec{X}(t))}{dt}=0,
\label{Eq:diss-drive}
\end{equation}
where we used $g,\Gamma,\sqrt{\Gamma},H$ to label the energy terms originated from driving, dissipation, noise, and the time-independent part of Eq.(\ref{eq:H}), respectively. For the time-independent Hamiltonian term, $\dot{E}_{H} =0$. If we now take the time average, the contribution from the noise $\overline{\dot{E}}_{\sqrt{\Gamma}}$ also vanishes. 
This means that, focusing only on the average values, the input power from the drive $\overline{\dot{E_g}}$ balances the dissipated energy $\overline{\dot{E}_\Gamma}$. 
The dynamics, in particular the Lyapunov exponent $\lambda$ for $\vec{X}(t)$, depends strongly on the 
Gaussian curvature of the $U(x,\chi)$ potential~\cite{Toda:1974,Brumer:1976,Pattanayak:1997b} 
along $\vec{X}(t)$, which can be sensitive to 
small changes in the steady-state mean ($\overline{H}$) and variance ($\Delta H$) of the total oscillator energy given by Eq.(\ref{eq:H}).

\section{Simulation results}

\begin{figure}%
    \centering
	 \subfloat[$\Gamma = 0.10$]{{\includegraphics[width=8cm]{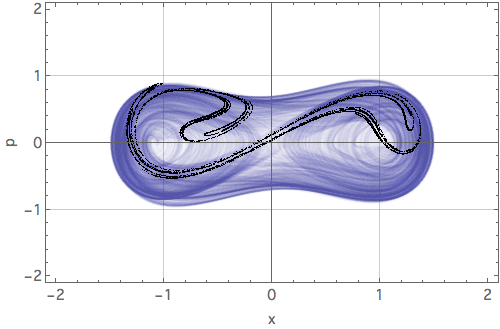} }}
    \qquad
     \subfloat[$\Gamma = 0.05$]{{\includegraphics[width=8cm]{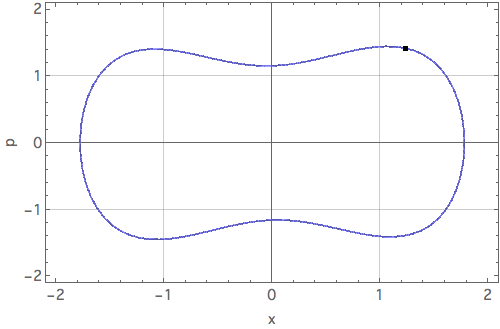} }}
    
    \caption{Phase space trajectories (blue) superimposed with Poincar\'e sections for the classical Duffing oscillator. Chaotic and regular behaviour are shown for $\Gamma=0.1$ (top) and $\Gamma=0.05$ (bottom), respectively.}%
    \label{fig:ClassicalPlots}%
\end{figure} 
\begin{figure}%
    \centering
    \subfloat[]{{\includegraphics[width=5cm]{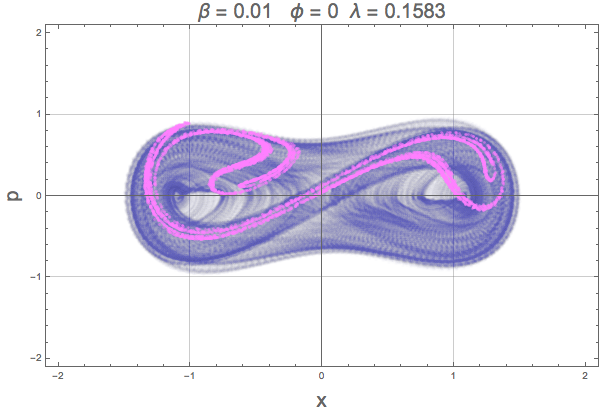} }}%
    \qquad
    \subfloat[]{{\includegraphics[width=5cm]{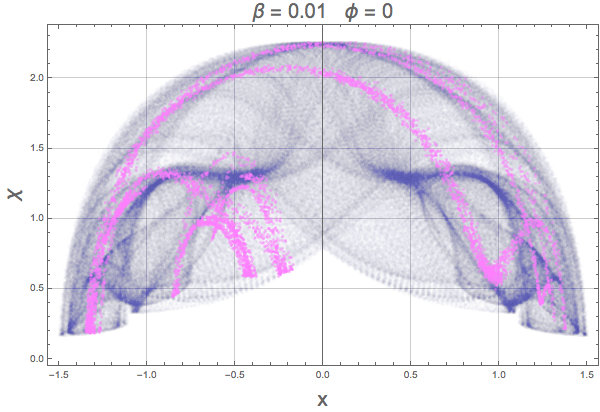} }}%
    \qquad
    \subfloat[]{{\includegraphics[width=5cm]{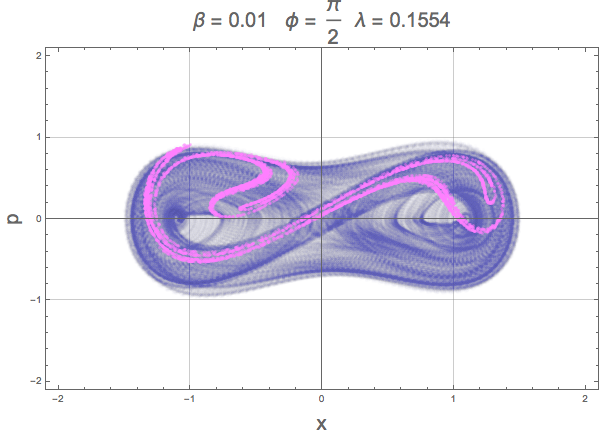} }}%
    \qquad
    \subfloat[]{{\includegraphics[width=5cm]{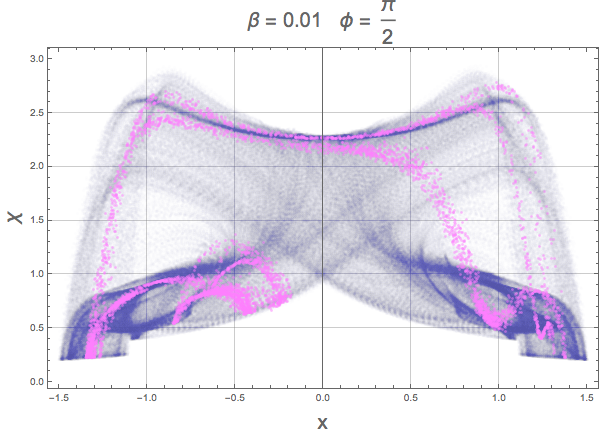} }}%
    \qquad
    \subfloat[]{{\includegraphics[width=5cm]{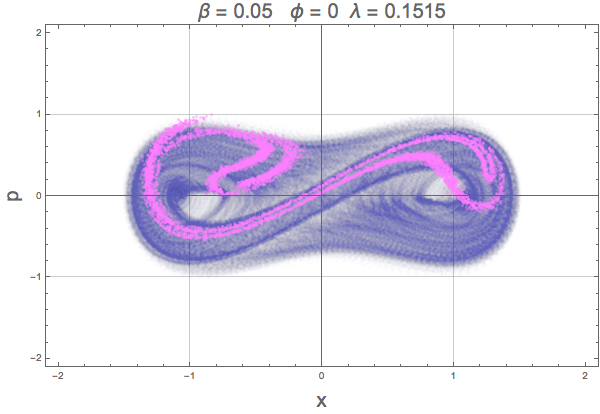} }}%
    \qquad
    \subfloat[]{{\includegraphics[width=5cm]{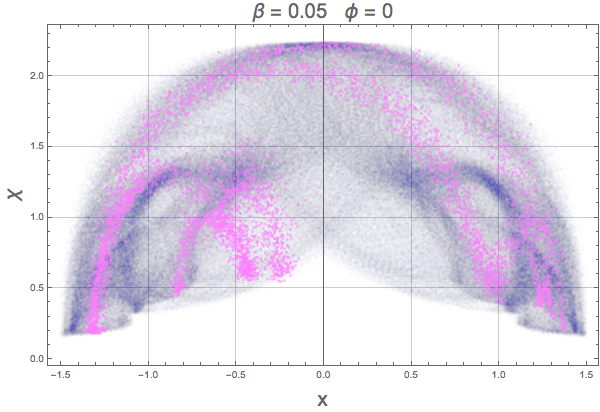} }}%
    \qquad
    \subfloat[]{{\includegraphics[width=5cm]{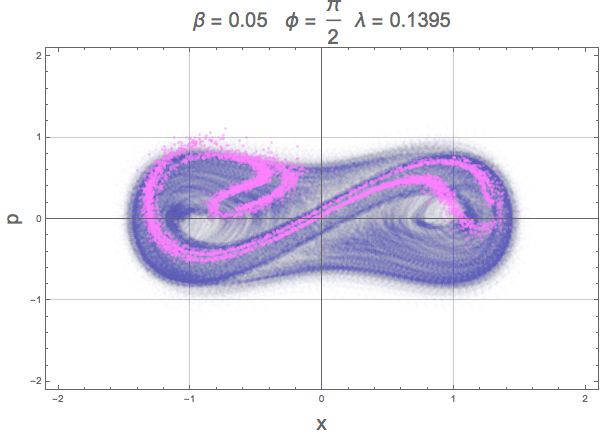} }}%
    \qquad
    \subfloat[]{{\includegraphics[width=5cm]{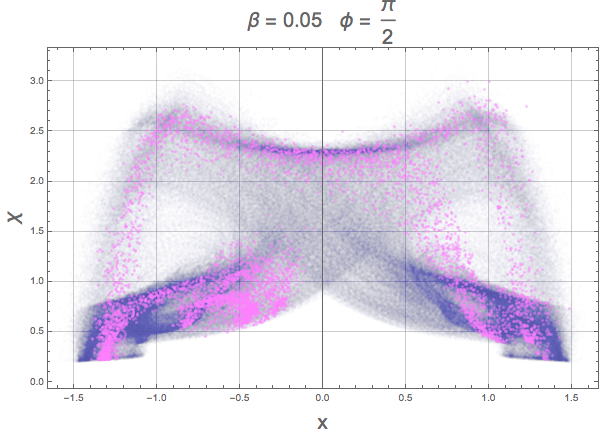} }}%
    \caption{$x$-$p$ (left) and $x$-$\chi$ (right) trajectories for $\Gamma = 0.1$. The values of $\beta$ were $0.01$ (a to d) and $0.05$ (e to h). For each case, the two measurement angles $\phi=0$ (a,b,e,f) and $\phi=\pi/2$ (c,d,g,h) were considered.}%
    \label{fig:Trajgam0.1}%
\end{figure}
\begin{figure}%
    \centering
    \subfloat[]{{\includegraphics[width=5cm]{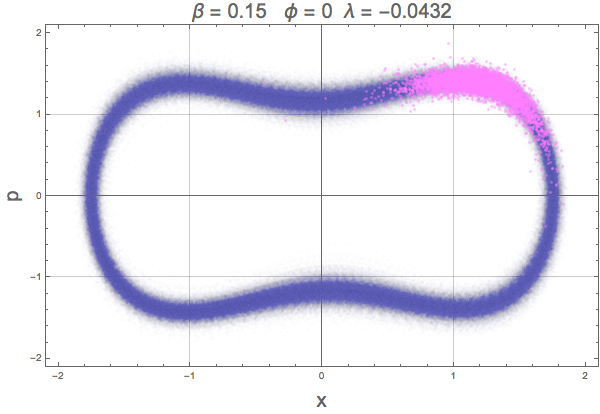} }}%
    \qquad
    \subfloat[]{{\includegraphics[width=5cm]{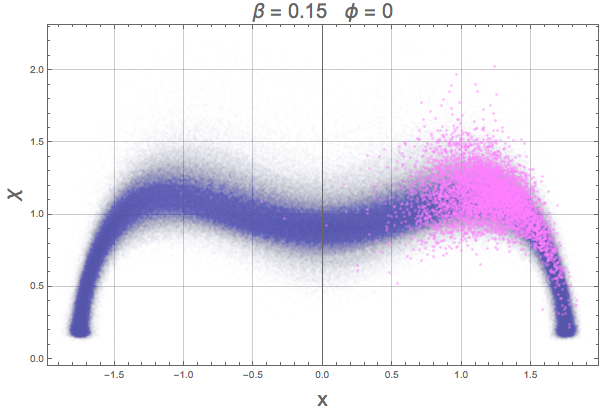} }}%
    \qquad
    \subfloat[]{{\includegraphics[width=5cm]{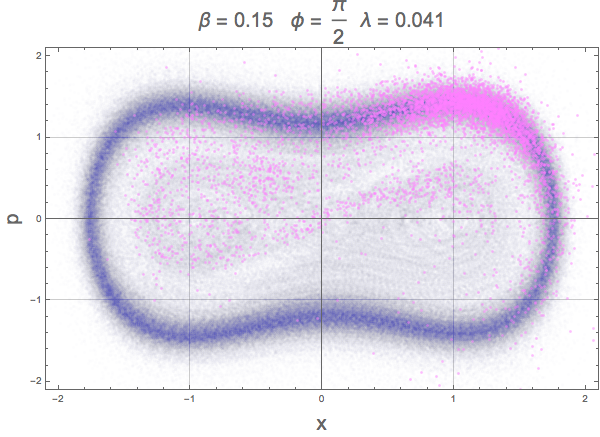} }}%
    \qquad
    \subfloat[]{{\includegraphics[width=5cm]{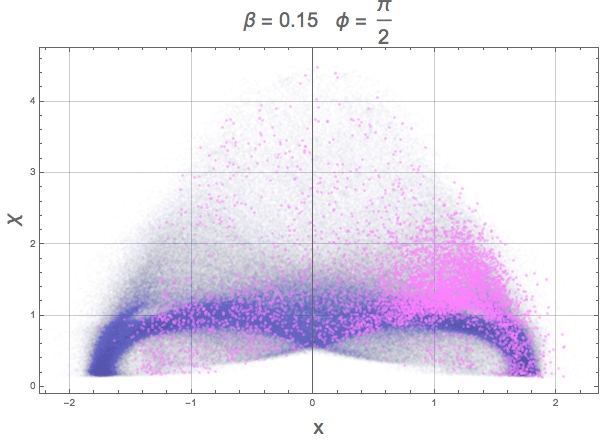} }}%
    \qquad
    \subfloat[]{{\includegraphics[width=5cm]{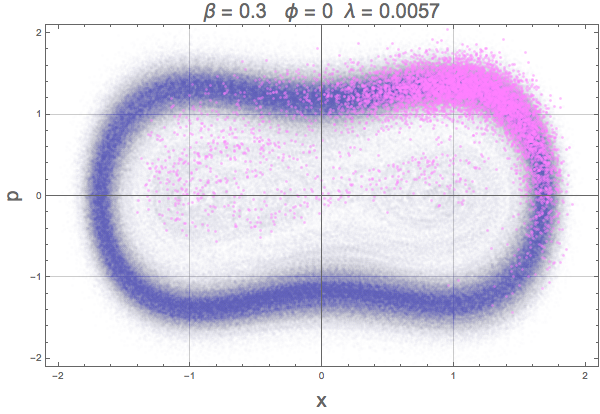} }}%
    \qquad
    \subfloat[]{{\includegraphics[width=5cm]{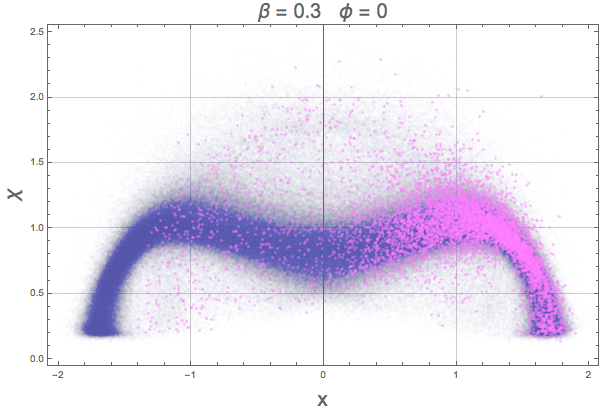} }}%
    \qquad
    \subfloat[]{{\includegraphics[width=5cm]{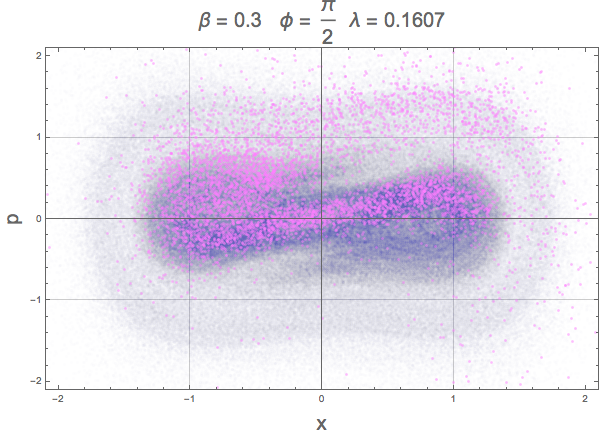} }}%
    \qquad
    \subfloat[]{{\includegraphics[width=5cm]{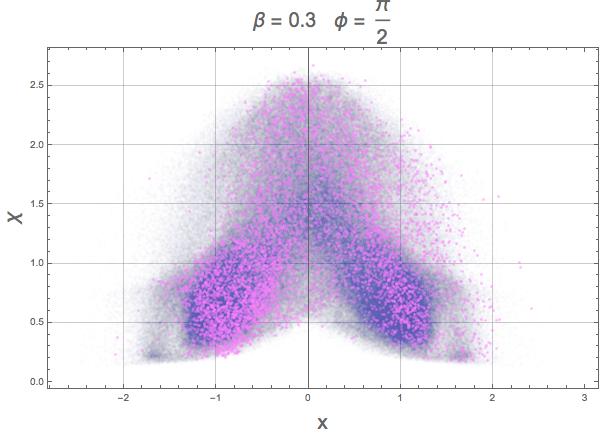} }}%
    \caption{$x$-$p$ (left) and $x$-$\chi$ (right) trajectories for $\Gamma = 0.05$ and two values of $\beta$: $0.15$ (a to d) and $0.3$ (e to h). For each case, the two measurement angles $\phi=0$ (a,b,e,f) and $\phi=\pi/2$ (c,d,g,h) were considered.}%
    \label{fig:Trajgam0.05}%
\end{figure}

Finally, we put together all the understanding developed in the previous sections to explain the semiclassical mechanism responsible for the reported~\cite{Eastman:2017} effects of measurement angle on quantum trajectory chaos. While for some parameter values the underlying phenomenon was shown to be purely quantum, for others, semiclassical effects seemed to play a role, but remained unexplained~\cite{Eastman:2017}. 

We consider the same two dissipative couplings $\Gamma_1 = 0.05$, $\Gamma_2 = 0.10$ previously studied in~\cite{Eastman:2017}. 
It is important to understand the difference in the classical limiting behavior at the two $\Gamma$ values. Consider the Poincar\'e sections (shown on top of corresponding trajectories) in the $(x,p)$ (classical) phase space in Fig.~(\ref{fig:ClassicalPlots}).
We notice that at low dissipation $\Gamma_1$ yields a simple inter-well periodic orbit that never goes inside the classical separatrix defined by the $H_1 (x,p)=0$ curve and has $\lambda < 0$. Hence the energy absorbed is dissipated exactly over a single period 
(although $\Delta H \neq 0$). However, at higher $\Gamma_2$, even though the orbit must dissipate what it absorbs on average since it stays confined in energy, the time-dependence of the dissipation term $\dot{E}_{\vec{F}}$ does not synchronize with the driving $\dot{E}_g$, such that the orbit wanders chaotically in a bounded energy range spanning the separatrix with $\lambda > 0$.

To understand the semiclassical behavior, for each $\Gamma$ we use both $\phi=0,\pi/2$ settings, and examine all 
these cases at two different length scales $\beta$. For each of these parameter combination, we show the 
Poincar\'e sections in $(x,p)$ as well as the $(x,\chi)$ space, the latter demonstrating how the range of $\chi$ affects classical behavior.

The first case analysed was for $\Gamma= 0.1$. Here we see that for both $\beta = 0.01$ and $\beta = 0.05$,  and irrespective of $\phi$, the quantum perturbations do not seem to visibly change the chaotic
 $(x,p)$ Poincar\'e sections. The $(x,\chi)$ Poincar\'e sections are very instructive, however. First note that the range of $\chi$ is essentially independent of $\beta$ for both $\phi$ values. On the other hand, the $\beta-$independent $\chi$ range for $\phi = \pi/2$ is much greater than for $\phi = 0$, consistent with our analysis of the role of the dissipative force for different measurement angles. As already observed in~\cite{Eastman:2017}, for this case, strong dependency of the Lyapunov exponent with the measurement angle is purely a quantum effect,  with little contribution of semiclassical origin. 

On the other hand, the case shown in Fig.~(\ref{fig:Trajgam0.05}) for $\Gamma_1$ is emblematic of the interplay between the two competing factors analysed in this paper: the coupling between centroid and spread variables, and measurement-dependent dissipation. At $\beta = 0.15$, the $\phi = 0$ case has smaller $\overline{U}_2$, $\overline{U}_{12}$ (visible in the range in $\chi$) than for $\phi = \pi/2$. Consistent with our previous discussion, for $\phi = 0$, the dissipative force pulls the system towards smaller values of $\chi$, leading, therefore, to the observed smaller values of  $\overline{U}_2$ and $\overline{U}_{12}$. For $\phi=\pi/2$, the dissipative force is not as effective in suppressing the effect of the nonlinear spread-centroid coupling, therefore the quantum corrections perturb the classical energy synchronization and induce chaos.
At $\beta = 0.3$, the semiclassical approximation is in principle not valid, but we find the same qualitative behavior with a full quantum simulation. Semiclassically, $\overline{U}_2$, $\overline{U}_{12}$ for $\phi = 0$ is smaller than for $\phi=\pi/2$. But the larger value of $\beta$ allows both angle settings to destroy the periodic motion although, again, chaos is stronger for $\phi=\pi/2$. It is worth noticing, from both the visual Poincar\'e sections as well as quantitatively from the $\lambda$ obtained, that $\beta =0.15$, $\phi =\pi/2$ shows larger $\overline{U}_2,\overline{U}_{12}$ values than for $\beta = 0.3$, $\phi = 0$ case such that it is effectively a more quantum system, and affects the classical motion to a greater extent. 

\section{Conclusion}

In closing, we have shown that a semi-classical nonlinear oscillator that is weakly monitored and coupled to the environment can be accurately understood as a classical centroid oscillator coupled to a `quantum' spread oscillator via a nonlinear $U_{12}$ coupling. We find that the the choice of measurement angle $\phi$ should be understood through its change on the dissipative measurement back-action that can dramatically alter how the nonlinear dynamics amplifies the size of $U_{12}$ to perturb the classical dynamics, sometimes substantially.

This leads to the remarkable observation that, comparing across all the parameter combinations presented, the measurement angle $\phi$ is more relevant than system scale $\beta$ in determining the dynamical regime of the system.

We are currently working on applications of these insights deep in the quantum regime where different mechanisms apply, as well as to adaptive control and quantum thermodynamics.

Acknowledgements:
All those at Carleton would like to thank Bruce Duffy for computational support, and AP would like to thank the Kolenkow-Reitz and the Towsley funds at Carleton College for support of students. AP and AC would like to thank the organizers of the Quantum Thermodynamics Conference 2018 in Santa Barbara for the excellent opportunity to learn and have conversations that partially led to this manuscript. AC also thanks AP's hospitality during his visits to Carleton College, where part of this work was developed. SG and JE gratefully acknowledge support by the Australian Research Council Centre of Excellence for Quantum Computation and Communication Technology (project number CE110001027).

\bibliographystyle{spmpsci}
\bibliography{allbib}

\begin{thebibliography}{10}
\providecommand{\url}[1]{{#1}}
\providecommand{\urlprefix}{URL }
\expandafter\ifx\csname urlstyle\endcsname\relax
  \providecommand{\doi}[1]{DOI~\discretionary{}{}{}#1}\else
  \providecommand{\doi}{DOI~\discretionary{}{}{}\begingroup
  \urlstyle{rm}\Url}\fi

\bibitem{Brumer:1976}
Brumer, P., Duff, J.W.: A variational equations approach to the onset of
  statistical intramolecular energy transfer.
\newblock The Journal of Chemical Physics \textbf{65}(9), 3566--3574 (1976).
\newblock \doi{10.1063/1.433586}.
\newblock \urlprefix\url{https://doi.org/10.1063/1.433586}

\bibitem{Brun:1996}
Brun, T.A., Percival, I.C., Schack, R.: Quantum chaos in open systems: a
  quantum state diffusion analysis.
\newblock Journal of Physics A: Mathematical and General \textbf{29}(9),
  2077--2090 (1996).
\newblock \urlprefix\url{http://stacks.iop.org/0305-4470/29/2077}

\bibitem{Eastman:2017}
Eastman, J.K., Hope, J.J., Carvalho, A.R.: Tuning quantum measurements to
  control chaos.
\newblock Scientific Reports \textbf{7}, 44,684 EP -- (2017).
\newblock \urlprefix\url{http://dx.doi.org/10.1038/srep44684}

\bibitem{Gorini:1976}
Gorini, V., Kossakowski, A., Sudarshan, E.C.G.: Completely positive dynamical
  semigroups of n-level systems.
\newblock J. Math. Phys. \textbf{17}, 821 (1976)

\bibitem{Halliwell:1995}
Halliwell, J., Zoupas, A.: Quantum state diffusion, density matrix
  diagonalization, and decoherent histories: A model.
\newblock Phys. Rev. D \textbf{52}, 7294--7307 (1995).
\newblock \doi{10.1103/PhysRevD.52.7294}.
\newblock \urlprefix\url{https://link.aps.org/doi/10.1103/PhysRevD.52.7294}

\bibitem{Kapulkin:2008}
Kapulkin, A., Pattanayak, A.K.: Nonmonotonicity in the quantum-classical
  transition: Chaos induced by quantum effects.
\newblock Physical Review Letters \textbf{101}(7), 074101 (2008).
\newblock \doi{10.1103/PhysRevLett.101.074101}.
\newblock \urlprefix\url{http://link.aps.org/abstract/PRL/v101/e074101}

\bibitem{Li:2012}
Li, Q., Kapulkin, A., Anderson, D., Tan, S.M., Pattanayak, A.K.: Experimental
  signatures of the quantum-classical transition in a nanomechanical oscillator
  modeled as a damped-driven double-well problem.
\newblock Physica Scripta \textbf{2012}(T151), 014,055 (2012).
\newblock \urlprefix\url{http://stacks.iop.org/1402-4896/2012/i=T151/a=014055}

\bibitem{Lindblad:1976}
Lindblad, G.: On the generators of quantum dynamical semigroups.
\newblock Math. Phys. \textbf{48}, 119 (1976)

\bibitem{Ota:2005}
Ota, Y., Ohba, I.: Crossover from classical to quantum behavior of the duffing
  oscillator through a pseudo-lyapunov-exponent.
\newblock Phys. Rev. E \textbf{71}, 015,201 (2005).
\newblock \doi{10.1103/PhysRevE.71.015201}.
\newblock \urlprefix\url{http://link.aps.org/doi/10.1103/PhysRevE.71.015201}

\bibitem{Pattanayak:1997}
Pattanayak, A.K., Brumer, P.: Chaos and lyapunov exponents in classical and
  quantal distribution dynamics.
\newblock Phys. Rev. E \textbf{56}, 5174--5177 (1997).
\newblock \doi{10.1103/PhysRevE.56.5174}.
\newblock \urlprefix\url{http://link.aps.org/doi/10.1103/PhysRevE.56.5174}

\bibitem{Pattanayak:1997b}
Pattanayak, A.K., Schieve, W.C.: Predicting two dimensional hamiltonian chaos.
\newblock Z. Naturforsch. \textbf{52a}, 34 (1997)

\bibitem{Pokharel:2018}
Pokharel, B., Misplon, M.Z.R., Lynn, W., Duggins, P., Hallman, K., Anderson,
  D., Kapulkin, A., Pattanayak, A.K.: Chaos and dynamical complexity in the
  quantum to classical transition.
\newblock Scientific Reports \textbf{8}(1), 2108 (2018).
\newblock \doi{10.1038/s41598-018-20507-w}.
\newblock \urlprefix\url{https://doi.org/10.1038/s41598-018-20507-w}

\bibitem{Rigo:1996}
Rigo, M., Gisin, N.: Unravellings of the master equation and the emergence of a
  classical world.
\newblock Quantum and Semiclassical Optics: Journal of the European Optical
  Society Part B \textbf{8}(1), 255 (1996).
\newblock \urlprefix\url{http://stacks.iop.org/1355-5111/8/i=1/a=018}

\bibitem{Toda:1974}
Toda, M.: Instability of trajectories of the lattice with cubic nonlinearity.
\newblock Physics Letters A \textbf{48}(5), 335 -- 336 (1974).
\newblock \doi{https://doi.org/10.1016/0375-9601(74)90454-X}.
\newblock
  \urlprefix\url{http://www.sciencedirect.com/science/article/pii/037596017490454X}

\bibitem{Wiseman:2001b}
Wiseman, H.M., Di\'osi, L.: Complete parameterization, and invariance, of
  diffusive quantum trajectories for markovian open systems.
\newblock Chemical Physics \textbf{268}(1-3), 91 -- 104 (2001).
\newblock \doi{DOI: 10.1016/S0301-0104(01)00296-8}.
\newblock
  \urlprefix\url{http://www.sciencedirect.com/science/article/pii/S0301010401002968}

\end{thebibliography}

\end{document}